# First Observations of Individual Molecular Clouds in the Irregular Galaxy NGC 6822

Christine D. Wilson

*Department of Physics and Astronomy, McMaster University, Hamilton, Ontario, Canada L8S 4M1*

## Abstract

Three molecular clouds have been mapped in the nearby metal-poor dwarf irregular galaxy NGC 6822 using the Owens Valley Millimeter-Wave Interferometer. These molecular clouds are similar to molecular clouds recently studied in the Small Magellanic Cloud, and are somewhat less massive on average than molecular clouds in Local Group spiral galaxies. Based on upper limits to the virial masses of the clouds, the CO-to-$H_2$ conversion factor in NGC 6822 is $< 2.2 \pm 1.3$ times larger than the Galactic value ($3 \times 10^{20}$ cm$^{-2}$ (K km s$^{-1}$)$^{-1}$). Conversion factors obtained for Galactic molecular clouds, NGC 6822, and the SMC are consistent with the variation in the CO-to-$H_2$ conversion factor being linearly proportional to the oxygen abundance of the galaxy. However, the uncertainties are sufficiently large that equal conversion factors in NGC 6822 and the Galaxy cannot be ruled out. The gas depletion time obtained for the HII region Hubble V is comparable to those of giant HII regions in M33. Thus high-efficiency star formation is not limited to the giant HII regions seen spiral galaxies, but can also occur in a small group of relatively low-mass (few $\times 10^4$ $M_\odot$) molecular clouds.

*Subject headings:* galaxies: individual (NGC 6822) – galaxies: irregular – galaxies: ISM – ISM: molecules – Local Group

## 1. Introduction

Irregular galaxies are important for studies of star formation, because they provide a relatively simple environment in which to study the star formation process, one that is free of dynamical driving effects such as density waves and nuclear bars. However, understanding star formation in irregular galaxies is hindered by difficulties in measuring the amount of molecular gas, the fuel for star formation. The weak CO lines observed in irregular galaxies (Combes 1986; Tacconi & Young 1987; Arnault et al. 1988) may indicate either a true deficiency of molecular gas (and hence an increased star formation efficiency) or a lower CO emissivity per mass of gas relative to molecular gas in spiral galaxies. The weak CO emission is commonly attributed to the low metallicities of these galaxies (Maloney & Black 1988; Elmegreen 1989), but there has been little good observational evidence to indicate exactly how important this effect might be.

Local Group galaxies are important for calibrating the effect of metallicity on the CO-to-$H_2$ conversion factor, because they are the only galaxies for which it is possible to resolve individual giant molecular clouds. The CO-to-$H_2$ conversion factor can then be measured directly by comparing masses estimated from the virial theorem with the CO luminosity of the cloud (Wilson & Scoville 1990). In these studies, it is important to obtain sufficient spatial resolution to resolve individual molecular clouds, which have diameters of 10 to 100 pc (Scoville & Sanders 1987), since the larger structures seen at lower resolution may not be gravitationally bound. Among dwarf irregulars in the Local Group, CO emission has now been detected in the Magellanic Clouds, NGC 6822, and IC 10. The Magellanic Clouds have been the target of a Key Project on the Swedish-ESO Submillimeter Telescope (Israel et al. 1993; Rubio, Lequeux, & Boulanger 1993a; Rubio et al. 1993b; Johansson 1991; Garay et al. 1993). The eleven molecular clouds mapped in the Small Magellanic Cloud (SMC) are smaller and less luminous in CO than Galactic giant molecular clouds (Rubio et al. 1993a). Despite its unusually strong CO lines (Becker 1990), studies of the molecular gas in IC 10 are hindered by the lack of a good distance determination. Although the CO lines in NGC 6822 are considerably weaker than in IC 10 (Elmegreen, Elmegreen & Morris 1980; Israel 1991; Wilson 1992a; Ohta et al. 1993), it should be possible to detect and resolve individual molecular clouds in this galaxy, provided the CO-to-$H_2$ conversion factor is not much larger than the Galactic value. NGC 6822 is actively forming stars, as evidenced by its bright HII regions (Hodge, Kennicutt, & Lee 1988) and OB associations (Hodge 1977; Wilson 1992b) and so should contain significant amounts of molecular gas. This letter reports the results from the first interferometric observations of individual molecular clouds in NGC 6822.

## 2. Observations and Analysis

One field in NGC 6822 centered on the giant HII region Hubble V, the strongest CO detection of Wilson (1992a), was observed in the CO J=1-0 line with the three-element Owens Valley Millimeter-Wave Interferometer during the 1991-92 observing season. The field center was $(\alpha(1950), \delta(1950)) = (19^h\ 42^m\ 03.9^s, -14^o\ 50'\ 06'')$. Due to the low elevation of the source ($38^o$ at transit), single-sideband system temperatures during the observations ranged from 600 to 1700 K. Absolute flux calibration was obtained by observations of Mars and Uranus and is estimated to be internally accurate to 10%.

The data analysis was carried out using the MIRIAD software package. The real and imaginary parts of the visibility data were clipped to a level of 21 Jy. The data were binned to a resolution of 1 MHz (2.6 km s$^{-1}$) with channels spaced every 0.5 MHz (1.3 km s$^{-1}$). Dirty maps were made of the individual channels using natural weighting to minimize the noise level in the maps. The rms noise in a 1 MHz channel map is 0.15 Jy beam$^{-1}$ and the beam size (full-width half-maximum) is $6.2 \times 11.1''$. Due to the low declination of the source, the beam has strong sidelobes at the 40% level separated from the main beam by about $45''$ (Fig. 1a). Inspection of the dirty channel maps revealed two regions of weak emission at the 4-5$\sigma$ level separated by about $45''$. Fig. 1b shows the dirty map integrated over the entire velocity width of the emission (5 MHz or 13 km s$^{-1}$).

To aid in identifying real sources in the dirty map, we obtained $^{12}$CO J=2-1 spectra of the four emission peaks during service observing with the James Clerk Maxwell Telescope (JCMT) on June 7, 1994. The beam width of the JCMT at 230 GHz is $22''$, sufficient to separate even the nearest emission peaks seen in Fig. 1b. The individual spectra shown in Fig. 2 have been smoothed to a resolution of 1.25 MHz (1.6 km s$^{-1}$) and a linear baseline has been subtracted. The total integration time is 10 minutes per position. Emission is detected at the 4-7$\sigma$ level from three of the positions, but is clearly *not* detected at the strongest emission peak in Fig. 1b. Based on the single dish spectra, the channel maps and integrated intensity map were cleaned using two clean boxes, one centered on the southern emission region and one on the north-eastern emission region (as shown in Fig. 1c). This procedure forces the cleaning algorithm to search for emission peaks only within the clean boxes, and to ignore emission in the rest of the field. A 1$\sigma$ flux cutoff was used in the cleaning process. The resulting cleaned integrated intensity map is shown in Fig. 1c. The fourth emission peak seen in the dirty map has completely disappeared, and thus has been fit quite well by the combination of the sidelobes from the other three clouds.

Three clouds were identified and their properties measured from the 1 MHz channel maps using the procedures described in Wilson & Scoville (1990). Table 1 gives the measured properties for the three clouds and their spectra are shown in Fig. 3. The integrated intensity and peak brightness temperature have been corrected for the response of the primary beam. A comparison of the total flux detected in the interferometer map with the single dish detection of Wilson (1992a) reveals that $46 \pm 14\%$ of the single dish flux has been detected in the map. Both virial and molecular masses were calculated for the clouds using the formulae given in Wilson & Scoville (1990). The distance to NGC 6822 has been measured to be 0.50 Mpc from near-infrared observations of Cepheid variables (McAlary et al. 1985). A value for the CO-to-H$_2$ conversion factor $\alpha_{N6822} = \alpha_{Gal} = (3 \pm 1) \times 10^{20}$ cm$^{-2}$ / (K km s$^{-1}$) (Strong *et al.* 1988; Scoville & Sanders 1987) was adopted initially. The diameter adopted for the virial mass calculation is $D_{pc} = 1.4\overline{D}_{FWHM} = 0.7(D_\alpha + D_\delta)$ (see Wilson & Scoville 1990 for details). Since none of the clouds are resolved in both dimensions, both the average diameters and virial masses are upper limits.

## 3. Discussion

The properties of the molecular clouds in NGC 6822 are very similar to those of molecular clouds in the SMC (Rubio et al. 1993ab), which have line widths of 3-6 km s$^{-1}$, diameters of 20-40 pc, virial masses of 2-9×10$^4$ M$_\odot$, and peak brightness temperatures of 0.5-2.4 K. The molecular clouds in NGC 6822 and the SMC are less massive on average than giant molecular clouds in the Milky Way (Scoville & Sanders 1987) or in M33 (Wilson & Scoville 1990). Fig. 4 compares the three molecular clouds in NGC 6822 with the size-line width relation obtained for nine clouds in M33. The agreement is reasonable given the upper limits to the cloud diameters, and suggests that the clouds in NGC 6822 are consistent with the size-line width relation determined for Galactic, M33, and SMC molecular clouds. With our limited data, we cannot determine whether NGC 6822 contains any substantially larger molecular clouds than the two studied here. However, the data for the SMC (Rubio et al. 1993a) combined with the difficulty in detecting CO in NGC 6822 at lower resolution (Wilson 1992a) suggest that the largest molecular clouds in these two irregular galaxies are probably substantially smaller than the clouds seen in spiral galaxies.

We can use the ratio of the virial and molecular masses to determine an upper limit to the ratio of the CO-to-H$_2$ conversion factors in NGC 6822 and the Galaxy. The molecular gas masses in Table 1 are systematically somewhat smaller than the virial masses. The ratios of the two mass estimates give values for $\alpha_{N6822}/\alpha_{Gal}$ of $< 3.8 \pm 2.1$ for MC1, $< 1.3 \pm 0.8$ for MC2, and $< 1.6 \pm 1.2$ for MC3, with an average value for all three clouds of $< 2.2 \pm 1.4$. This CO-to-H$_2$ conversion factor is somewhat smaller than the conversion factor recently determined for 11 molecular clouds in the Small Magellanic Cloud ($\sim$ 4 times the Galactic value, Rubio et al. 1993a). However, the oxygen abundance in the HII region Hubble V is $12 + log(O/H) = 8.16 \pm 0.06$ (Lequeux et al. 1979; Pagel, Edmunds, & Smith 1980; Skillman, Terlevich, & Melnick 1989), 2.3 times smaller than Orion ($12 + log(O/H) = 8.52$, Peimbert & Torres-Peimbert 1977) and 1.5 times larger than the SMC ($12 + log(O/H) = 7.98$, Pagel et al. 1978). It is striking that the relative conversion factors in these three galaxies are consistent with a CO-to-H$_2$ conversion factor that depends linearly on the oxygen abundance of the galaxy, a trend which is particularly evident when we compare abundances relative to the Orion molecular cloud rather than to the solar oxygen abundance ($12 + log(O/H) = 8.92$, Lambert 1978). However, although the molecular clouds in NGC 6822 appear underluminous in CO by about a factor of two relative to Galactic molecular clouds, the uncertainties in the measurement are sufficiently large that we cannot rule out the possibility that NGC 6822 and the Galaxy have equal CO-to-H$_2$ conversion factors. Thus these data provide only weak confirmation of the recent result for SMC molecular clouds, that clouds in a low-metallicity environment have relatively weak CO luminosities (Rubio et al. 1993a).

The integrated CO map is overlaid on a blue optical image from Wilson (1992b) in Fig. 5. MC1 and MC2 coincide quite well with the position of the HII region Hubble

V, the brightest HII region in NGC 6822 (Kennicutt 1988). Combining the upper limit to the CO-to-$H_2$ conversion factor obtained here with the single dish flux measured by Wilson (1992a) gives an upper limit to the total $H_2$ gas mass associated with Hubble V of $< 1.9 \times 10^5$ $M_\odot$. The star formation rate for Hubble V was estimated from published data to be $4 \times 10^{-3}$ $M_\odot$ yr$^{-1}$ (Wilson 1992a), which can be combined with the $H_2$ mass to give a gas depletion time of $< 5 \times 10^7$ yr. This short gas depletion time is very similar to those measured for the two largest HII regions of M33, which have star formation rates roughly an order of magnitude larger than Hubble V (Wilson & Scoville 1992). Thus whatever mechanism is responsible for the high star formation rates and efficiencies found in giant HII regions, it must be able to operate not only in the massive agglomerations of molecular gas found in spiral galaxies, but also in a collection of a few low-mass molecular clouds.

## 4. Conclusions

Three molecular clouds in the Local Group dwarf irregular galaxy NGC 6822 have been mapped with high spatial resolution using the Owens Valley interferometer. These clouds are very similar to molecular clouds recently identified in the Small Magellanic Cloud by Rubio et al. (1993ab), and are on average somewhat smaller than the molecular clouds seen in Local Group spiral galaxies. The clouds appear to be somewhat underluminous in CO for their virial mass; based on upper limits to the virial masses of the three clouds, the CO-to-$H_2$ conversion factor in NGC 6822 is estimated to be $< 2.2 \pm 1.4$ times larger than in the Milky Way or M33. This value is intermediate between the Galactic value and the conversion factor obtained recently for clouds in the SMC (Rubio et al. 1993a), and is consistent with the variation in the CO-to-$H_2$ conversion factor being linearly proportional to the oxygen abundance. However, the uncertainties in the measurement are sufficiently large that equal conversion factors in NGC 6822 and the Galaxy cannot be ruled out. Using this new upper limit for the conversion factor, the gas depletion timescale in the HII region Hubble V is found to be comparable to those of HII regions in M33 that are ten times more massive. Thus high-efficiency star formation is not limited to the giant HII regions seen spiral galaxies, but can also occur in a collection of a few low-mass (few $\times 10^4$ $M_\odot$) molecular clouds.

## Acknowledgements

I thank Karen Bakker for help with the re-calibration of the data. This research was partially supported by a Women's Faculty Award and a research grant from NSERC (Canada).

**TABLE 1**
Properties of Molecular Clouds in NGC 6822

| Cloud | MC1 | MC2 | MC3 |
|---|---|---|---|
| CO J=1-0 line | | | |
| $\alpha(1950)$ ($^h$ $^m$ $^s$) | 19 42 4.4 | 19 42 3.3 | 19 42 4.9 |
| $\delta(1950)$ ($^\circ$ $'$ $''$) | -14 50 31 | -14 50 27 | -14 49 54 |
| $V_{peak}$ (km s$^{-1}$) | -41.3 | -41.3 | -40.0 |
| $V_{FWHM}$ (km s$^{-1}$)$^a$ | 4.7 | 3.6 | 2.9 |
| $T_B$ (K) | 1.3±0.2 | 1.3±0.2 | 1.1±0.2 |
| $S_{CO}$ (Jy km s$^{-1}$)$^a$ | 2.9 | 12.3 | 3.5 |
| $D_\alpha \times D_\delta$ (pc)$^{a,b}$ | 12× < 18 | 52× < 18 | < 10 × 28 |
| $M_{vir}$ (10$^4$ M$_\odot$) | < 4.6±2.3 | < 6.3 ± 3.3 | < 2.2 ± 1.5 |
| $M_{mol}$ (10$^4$ M$_\odot$)$^a$ | 1.2 | 5.0 | 1.4 |
| CO J=2-1 line | | | |
| $T_{peak}$ (K)$^c$ | 0.31 ± 0.06 | 0.44 ± 0.06 | 0.22 ± 0.05 |
| $S_{CO}$ (K km s$^{-1}$)$^c$ | 1.46 ± 0.32 | 2.38 ± 0.33 | 0.89 ± 0.25 |

A distance to NGC 6822 of 0.50 Mpc is assumed throughout.
$^a$ The typical errors are: $D$ ±5 pc, $\Delta V_{FWHM}$ ±1.3 km s$^{-1}$, $S_{CO}$ ±20%, $M_{mol}$ ±30%
$^b$ $D_\alpha$ and $D_\delta$ are the deconvolved full-width half-maximum diameters in the right ascension and declination directions.
$^c$ Single-dish data are on the main-beam temperature scale, $T_{mb} = T_A^*/\eta_{mb}$, $\eta_{mb} = 0.72$.

## Figure Captions

**Fig. 1** – a) The synthesized beam of the NGC 6822 observations; the contour intervals are (-2,-1,1,2,3,...,10)×10% of the peak. The coordinates of the field center are $19^h\ 42^m\ 03.9^s$, $-14^o\ 50'\ 06''$ (1950). b) The dirty map of the NGC 6822 data. The emission is integrated over 5 MHz (13 km s$^{-1}$) centered on $V_{lsr} = -39.4$ km s$^{-1}$. The contour intervals are (-4,-3,-2,2,3,4,5,6,7)×0.053 Jy beam$^{-1}$ ($1\sigma$). The half-power beam width of the primary beam (66'') is indicated. c) The cleaned integrated intensity map obtained using the clean boxes indicated. The contour levels are the same as Fig. 1b. The map has not been corrected for attenuation due to the primary beam.

**Fig. 2** – CO J=2-1 single-dish spectra for the four positions showing J=1-0 emission in the dirty interferometer map. Each spectrum is labeled with its coordinates (1950.0).

**Fig. 3** – CO spectra for the three molecular clouds in NGC 6822, integrated over the area of the cloud. Each channel is 1 MHz wide and the channels are separated by 0.5 MHz, so adjacent channels are not independent. The fluxes have been corrected for the response of the primary beam. a) MC1. b) MC2. c) MC3.

**Fig. 4** – The three clouds identified in NGC 6822 are plotted on the size-line width relation obtained for molecular clouds in M33 by Wilson & Scoville (1990). The clouds in NGC 6822 are indicated by large open circles and the M33 clouds by small filled circles. The curved line is a fit to the nine molecular clouds in M33. Note that the diameters of the NGC 6822 clouds are upper limits.

**Fig. 5** – The integrated intensity map of the emission in NGC 6822 is overlaid on a blue optical image. The bright concentration of stars is the OB association for the HII region Hubble V (Hodge 1977; Wilson 1992b). The contour intervals are (-4,-3,-2,2,3,4,5)×0.053 Jy beam$^{-1}$ ($1\sigma$). The map has not been corrected for attenuation due to the primary beam.

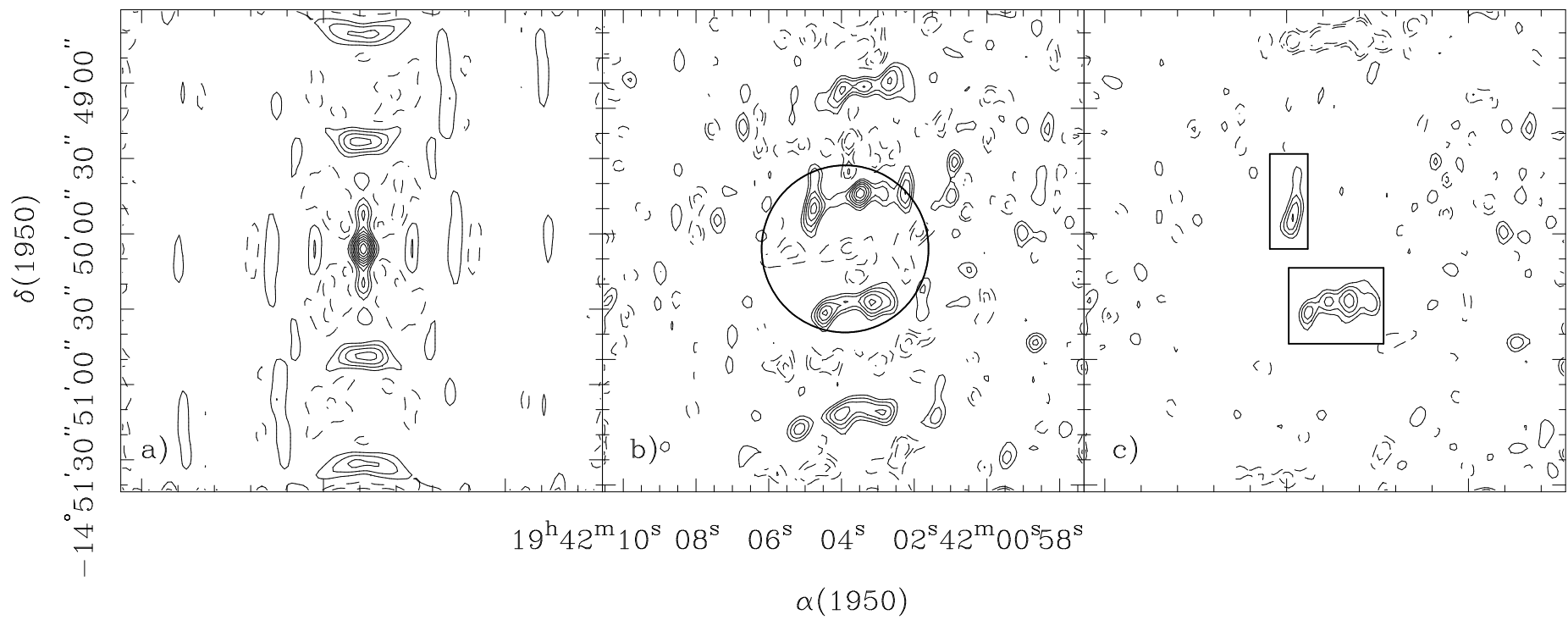

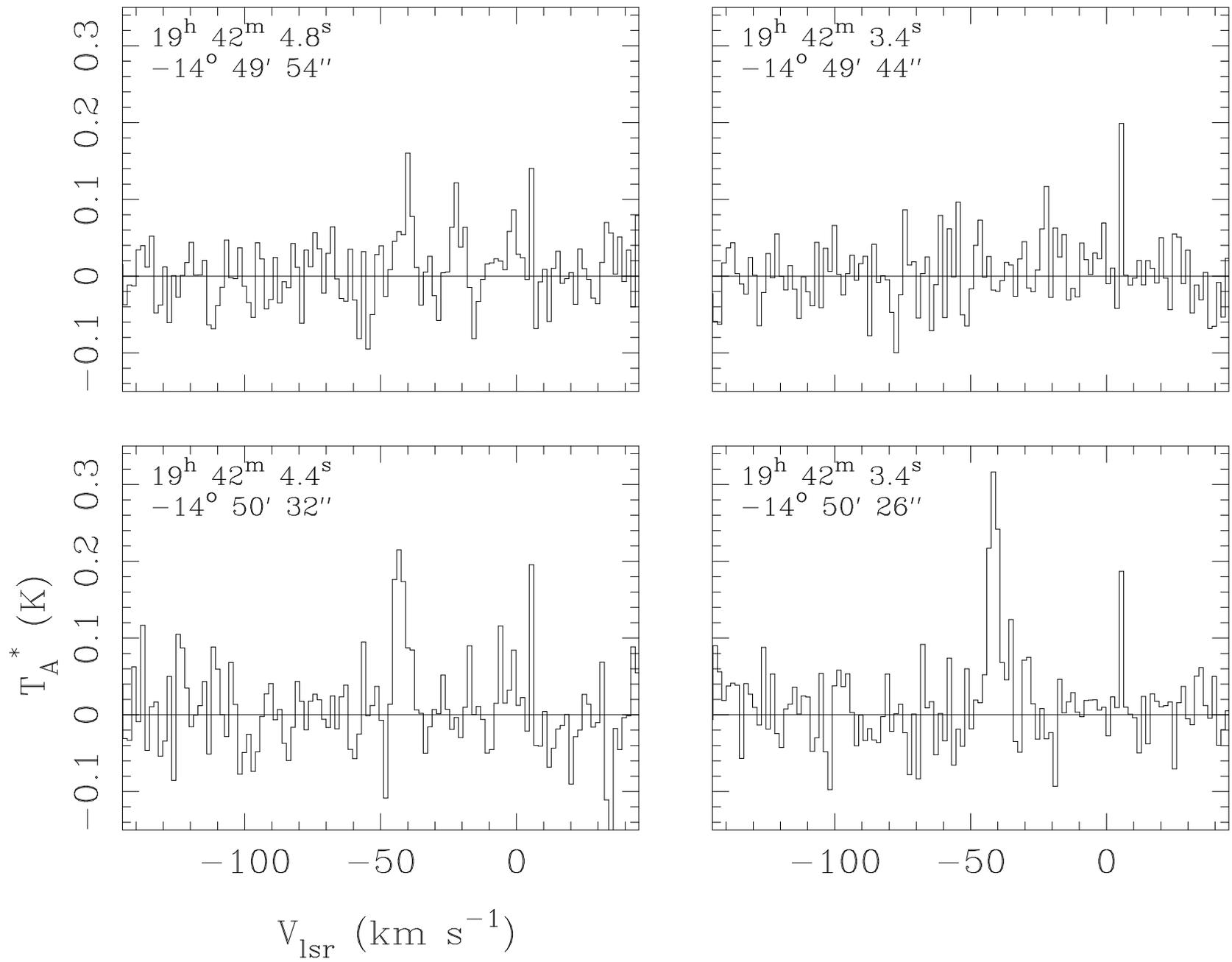

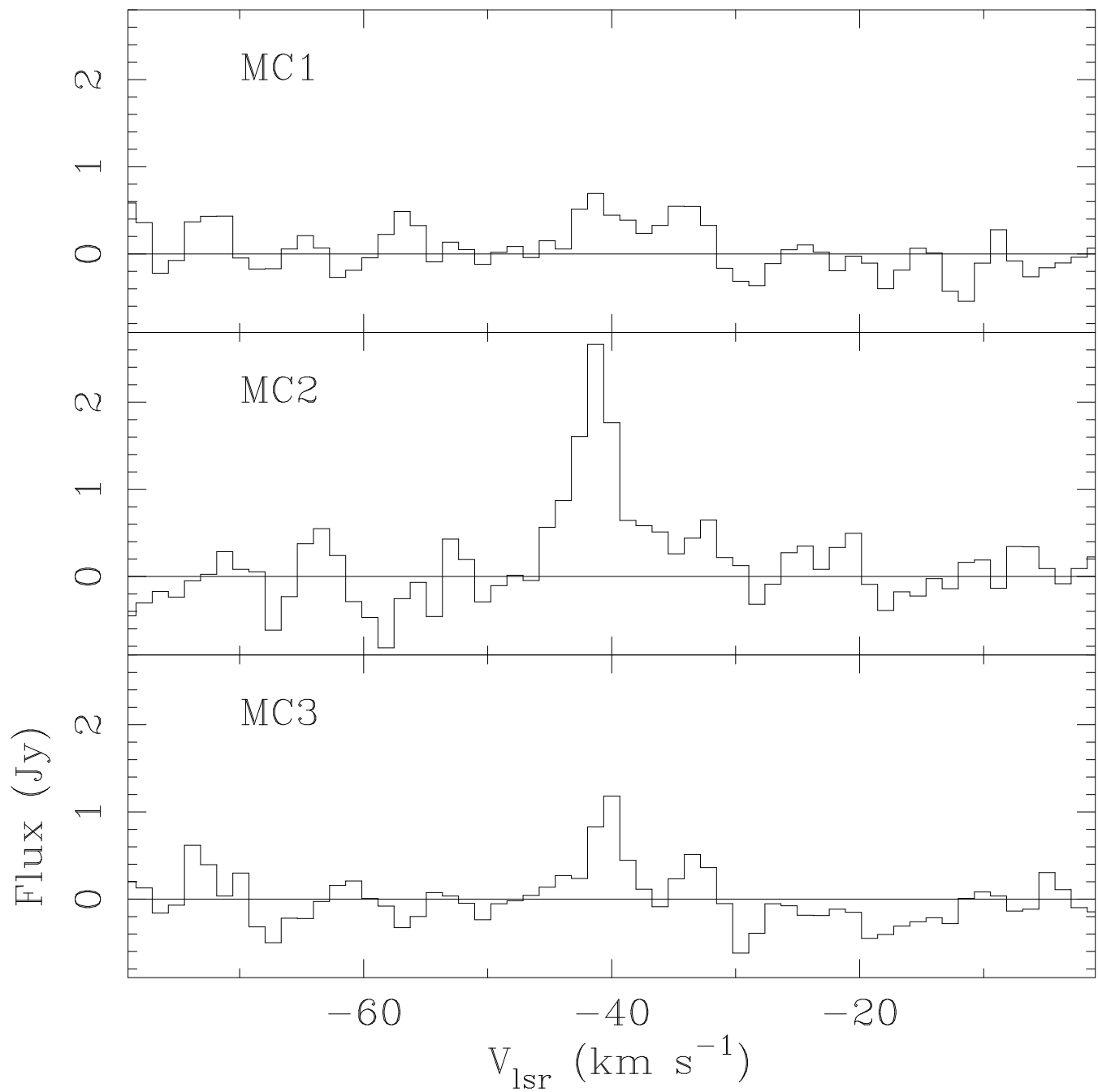

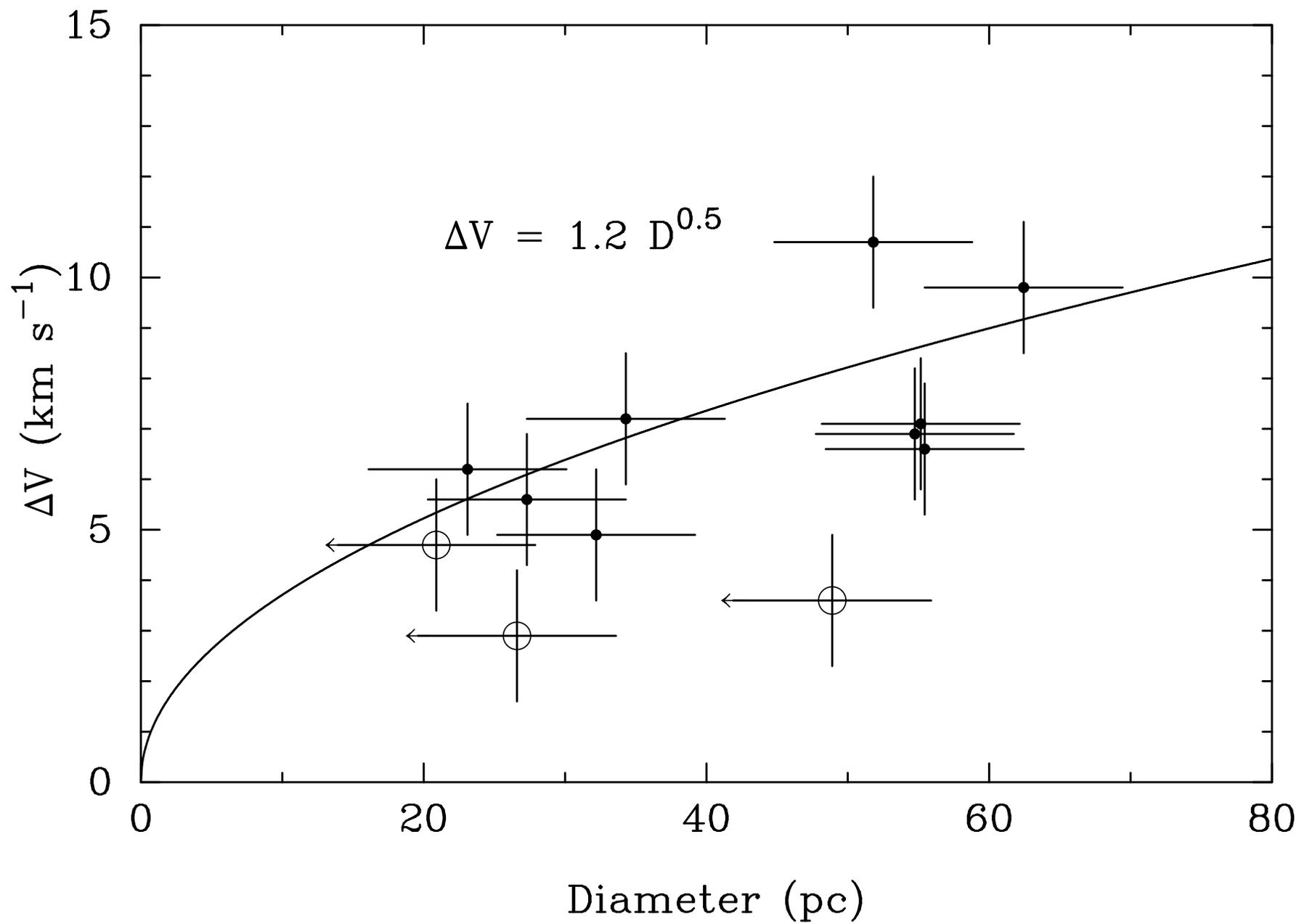

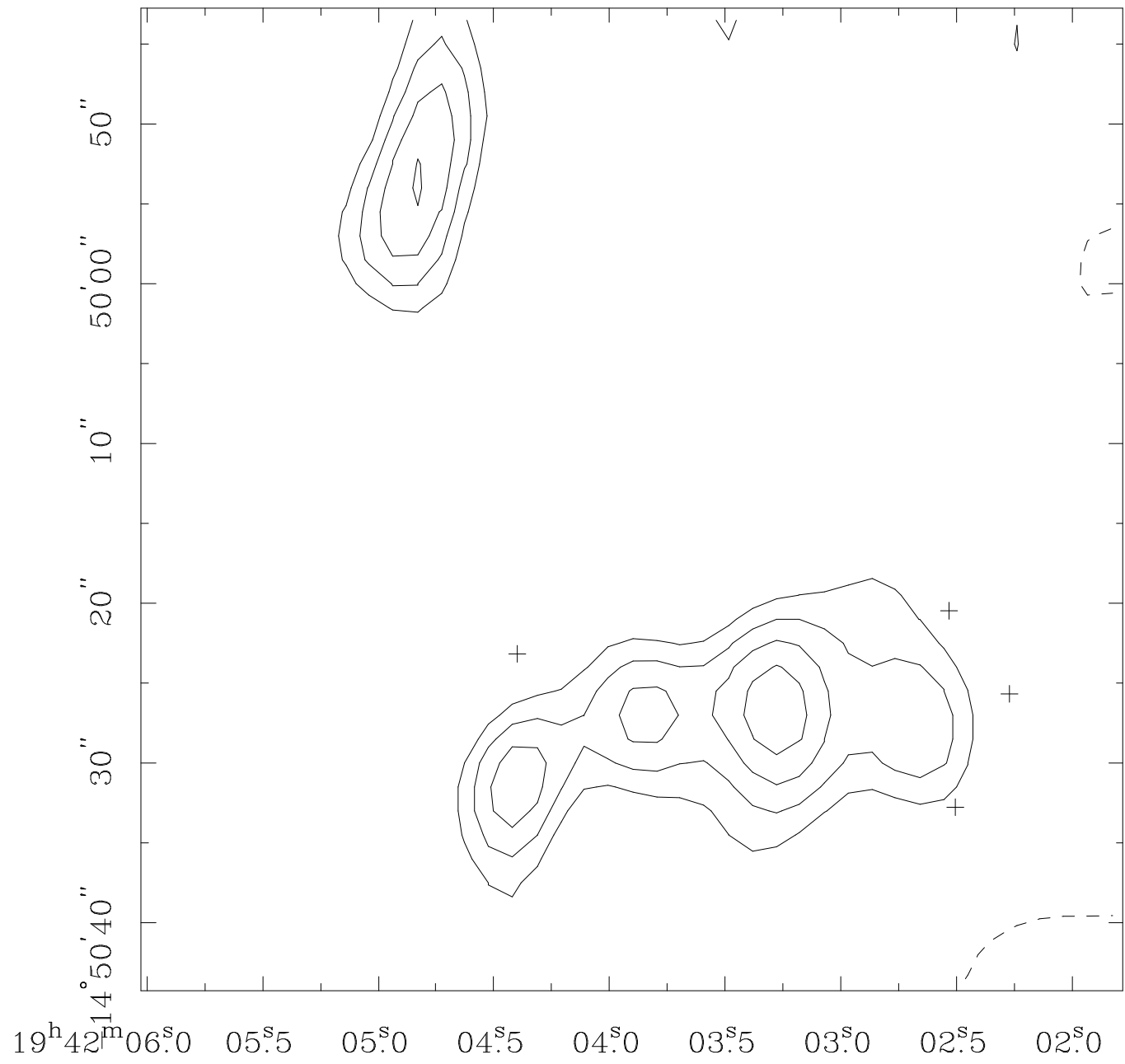